\documentclass[11pt,a4paper]{article}

\usepackage{amsxtra}
\usepackage{amsfonts}
\usepackage{amsmath,amssymb,stmaryrd} 
\usepackage[english]{babel} 
\usepackage{graphicx} 
\usepackage{mathrsfs} 
\usepackage{layout}
\usepackage{color}

\newtheorem{theo}{Theorem}
\newtheorem{prop}[theo]{Proposition}
\newtheorem{lem}[theo]{Lemma}
\newtheorem{cor}[theo]{Corollary}
\newtheorem{rem}[theo]{Remark}
\newtheorem{defin}[theo]{Definition}

\newcommand{\beqn}{\begin{equation}}
\newcommand{\eeqn}{\end{equation}}
\newcommand{\bear}{\begin{eqnarray}}
\newcommand{\eear}{\end{eqnarray}}
\newcommand{\bean}{\begin{eqnarray*}}
\newcommand{\eean}{\end{eqnarray*}}

\newcommand{\bml}{\begin{gather}} \newcommand{\eml}{\end{gather}}

\newcommand\qed{\hfill$\sqcap\kern-7.5pt\hbox{$\sqcup$}$}
\newcommand{\R}{{\mathbb R}}

\newcommand{\N}{{\mathbb N}}

\newcommand{\pa}{\partial}

\begin{document}
 
\title{Propagation of Chaos in a Coagulation Model}
\maketitle

\begin{center}
M. Escobedo\footnotemark[1]$^,$\footnotemark[2] and F. Pezzotti \footnotemark[3]
\end{center}
\footnotetext[1]{Departamento de Matem\'aticas, Universidad del
Pa{\'\i}s Vasco, Apartado 644, E--48080 Bilbao, Spain.  E-mail: {\tt
miguel.escobedo@ehu.es}} 
\footnotetext[2]{Basque Center for Applied Mathematics (BCAM), 
Bizkaia Technology Park, Building 500, E--48160 Derio, Spain. }
\footnotetext[3]{Departamento de Matem\'aticas, Universidad del
Pa{\'\i}s Vasco, Apartado 644, E--48080 Bilbao, Spain. E-mail: {\tt
federica.pezzotti@ehu.es}}

\noindent\textbf{Abstract.} A deterministic coalescing dynamics with constant rate for a particle system  in a finite volume with a fixed initial
number of particles is considered. It is shown that, in the thermodynamic limit, with the constraint of fixed
density, the corresponding coagulation equation is recovered and global in time 
propagation of chaos holds.


\noindent

\section{Introduction}
\label{Introduction}
\setcounter{equation}{0}
\setcounter{theo}{0}
Coagulation equations are widespread models describing large sets of coalescing particles. It is argued in such descriptions that the coagulation rate  depends on the masses of the particles. In the spatially homogeneous models it is further assumed that spatial fluctuations in the mass density are negligible. In the first example of this sort, proposed by Smoluchowski \cite{1916ZPhy...17..557S}, particles of radius $r$ moving by Brownian motion with variance proportional to $1/r$  meet at rate proportional to $(r_1+r_2)(r_1^{-1}+r_2^{-1})$. The concentration $f(t,m)$ of particles of mass $m=1, 2, 3,  \cdots$ at time $t$ satisfies the equation:
\bean
&&\frac{\partial f}{\partial t}(t, m)=\frac{1}{2}\sum_{k=1}^{m-1}K(m-k, k)f(t, m-k)f(t, k)-f(t, m)\sum_{k=1}^\infty K(m, k)f(t, k)\\
&&K(m, k)=\left(m^{1/3}+k^{1/3} \right)\left( m^{-1/3}+k^{-1/3}\right)
\eean
If masses are considered to take continuous positive values we are generally led to the continuous version:
\bear
\label{S1ECoagEqu}
\frac{\partial f}{\partial t}(t, m)=\frac{1}{2}\!\int_0^m K(m-m', m')f(t, m-m')f(t, m')dm'-f(t, m)\!\int_0^\infty  K(m, m')f(t, m')dm'
\eear

Different coagulation kernels $K(m, m')$ may be considered, in particular constant, additive and multiplicative kernels have been widely studied. 
The concentration $f(t,m)$ is defined as the average number of clusters of mass $m$ per unit volume at time $t$ in the discrete case,
and as the average number of clusters of mass in $[m, m+dm]$ per unit volume at time $t$ in the continuous case.
Our purpose is to deduce the continuous coagulation equation as the limit of equations  describing a coalescing system of
finite (large) number of particles   in a  finite spatial volume by means of  PDE arguments.  

To this end  we consider, in a given  volume $V$, a system of particles whose number changes in time due to the coagulation process.  At any time $t\ge 0$, given  any $N\in \N^\ast $ and $(m_1, m_2, \cdots, m_N)\in  (\R^+)^N $ we consider the  mass distribution functions $P_{N} (t, m_1,..., m_N)$. Each of these functions describe the state of  a system constituted by $N$ particles in a volume $V$ where, at time $t$, one particle has mass between $m_1$ and $m_1 + dm_1$, one has mass between $m_2$ and $m_2 + dm_2$ and so on. We may then define the probability to have $N$ particles in the volume $V$ at time $t=0$  as the following:
\bean
P(0, N)=\frac{1}{N!}\int_0^\infty\cdots\int_0^\infty P^0_{N} (m_1,..., m_N)dm_1\cdots dm_N
\eean
and
\bean
\sum_{N=1}^\infty P(0, N)=1
\eean
Given any $N_0\in \N^\ast$ we consider initial data satisfying the following conditions:

\begin{equation}
\label{initialdatum}
\left\{
\begin{array}{l}
(i)\,\,\,P^0_{N} (m_1,..., m_N)=0, \,\,\,\hbox{for all}\,\,N\not = N_0,\\ \\
(ii)\,\,\,P^0_{N_0} (m_1,...,m_{N_0})=(N_0)!f_0(m_1)\otimes\cdots\otimes f_0(m_{N_0}), \,\, m_i \in  (0,+\infty), \,\,\,i=1, 2, \dots N_0\\ \\
\hskip 0.7cm  \hbox{where:}\quad  f_0\ge 0,\,\, \displaystyle{\int_0^\infty f_0(m)\, dm=1}.
\end{array}
\right.
\end{equation}
Condition $(i)$ expresses that at $t=0$ the system has exactly $N_0$ particles.  It is easily seen that conditions $(i)$ and $(ii)$ imply  $P(0,N)=\delta(N-N_0)$. The average number of particles is then:
\bean
\sum_{N=1}^\infty N P(N, 0)=N_0.
\eean
as expected. 

The starting point of our analysis are the evolution equations satisfied by the  mass distribution functions $P_{N} (t, m_1,..., m_N)$ throughout the coagulation process. These equations had been obtained in \cite{MR0223151} and later in \cite{1972JAtS...29.1496G}, \cite{Tanaka1994404} and  \cite{Lushnikov1978276}.  It is then possible to deduce a  system  of equations for the correlation functions:
\begin{eqnarray*}
f_j(m_1,\dots, m_j,t) &=& \sum_{N=j}^\infty \frac{1}{(N-j)!}\int_0^{\infty} d m_{j+1}\dots \int_0^{\infty} d m_{N}\  P_{N}(m_1, \dots, m_N,t).
\end{eqnarray*}
where $j\in \N^\ast$. Notice that for $j=1$ we have
\bean
\overline N (t)&=&\sum_{N=1}^\infty NP(t, N)=\sum_{N=1}^\infty\frac{1}{(N-1)!}\int_0^\infty\cdots \int_0^\infty P_N(t, m_1, m_2, \cdots, m_N)\, dm_1\cdots dm_N\\
&=&\int_0^\infty f_1(t, m_1)\, dm_1
\eean
and therefore $f_1(t)$ is the density function associated to the mean number of particles at time $t$.

Consider now the following rescaled functions:
\bear
\label{S1Erescaled}
f_j^V(t, m_1, \cdots, m_j)=\frac{f_j(t, m_1, \cdots, m_j)}{V^j}.
\eear
We have, for $j=1$: 
\bean
&&\int_0^\infty f_1^V(t, m_1)dm_1=\frac{\overline {N(t)}}{V}= \hbox{mean number of particles per unit volume at time }\,\,t\ge 0.
\eean
We then study the limit of these rescaled  functions as
\bear
\label{S1Elimit}
N_0\to +\infty,\,\,V\to +\infty,\,\,\, \frac{N_0}{V}\to \rho _0.
\eear
By  (\ref{initialdatum}), we have
\bear
\label{initialdatumdos}
f_j^V(0, m_1, \cdots, m_j)=\frac {N_0(N_0-1)\dots (N_0-j+1)} {V^j}f_0^{\otimes\, j}(m_1, \cdots, m_j)
\eear and  it easily follows that:
\bear
\label{chaosmolecular}
\forall j\in \N^\ast:\,\,\, \lim_{\substack{V, \,N_0\to +\infty\\ \frac{N_0}{V}\to \rho_0}}||f_j^V(0)-(\rho_0)^j f_0^{\otimes j}||_{L^1((\R^+)^j)}=0.
\eear
The main result of the paper is the following.

\begin{theo}
\label{MAINTh} Suppose that the coagulation kernel $K$  is constant.
Let $\{f_j^V(t)\}_{j\in\mathbb N^*}$ be the sequence of functions defined in (\ref{S1Erescaled}) with initial data (\ref{initialdatumdos}). Then, 
\bear
\label{S1EMainThE1}
\forall \ \ j\ge 1:\qquad \lim_{\substack{V, \,N_0\to +\infty\\ \frac{N_0}{V}\to \rho_0}} ||f_j^V(t)-
f(t)^{\otimes j}||_{L^1((\R^+)^j)}=0
\eear
where $f$ is the unique solution in $\mathbf C([0, +\infty); L^1(\R^+))$ of the coagulation equation (\ref{S1ECoagEqu}) with initial datum $\rho_0 f_0$.
\end{theo}
The convergence result (\ref{S1EMainThE1}) under the hypothesis (\ref{initialdatumdos}) for the initial data is usually known as  propagation of chaos.
By the initial condition (\ref{initialdatumdos}) the particles are identically and independently distributed at time $t=0$. The conclusion (\ref{S1EMainThE1}) means that at any time $t>0$, the system of particles has still that property but only asymptotically in the limit  (\ref{S1Elimit}) (cf. \cite{MR2656357} and references therein).

The suitable way to describe the finite system of coalescing particles is the stochastic Markov process  called Marcus-Lushnikov process. This has been introduced in \cite{MR0223151} and later in \cite{1972JAtS...29.1496G}, \cite{Tanaka1994404} and  \cite{Lushnikov1978276}. In these four references the authors obtain and pay special attention to the evolution equation satisfied by the probability of the stochastic process, called joint frequency or mass  distribution function.  They deduce from that equation the system of equations satisfied by the moments of the mass distribution function.  The conditions under which this system may be approximated by the coagulation equation are discussed.  It is seen in particular  that this depends on the no correlation between the numbers of particles of different mass (cf. also \cite{1967JAtS...24..221S} for the continuous case). 

In   \cite{MR1631473}, \cite{springerlink:10.1023/A:1004640317274}, \cite{0944.60082}, \cite{Nicolas} the authors consider the  Marcus-Lushnikov process itself.  The weak convergence of suitably rescaled  versions of these processes, called stochastic coalescents,  towards the solutions of the coagulation equation is proved under different conditions on the coagulation kernel $K$,  both in the discrete and the continuous case.  

We believe nevertheless that  the questions raised in 
\cite{MR0223151}, \cite{1972JAtS...29.1496G}, \cite{Tanaka1994404} and  \cite{Lushnikov1978276} about the system of equations satisfied by the moments are of interest.
In particular, to study the  case of continuous masses  via the equations for the mass distribution functions by means of PDE arguments is worthwhile, for itself and for further work \cite{PE}.  That leads very naturally to a set of equations for the correlation functions  $f_j$'s  that, as we have already said, has been discussed to some extent  in  \cite{MR0223151}, \cite{1972JAtS...29.1496G}, \cite{Tanaka1994404} and  \cite{Lushnikov1978276}. The system of rescaled correlation functions $f_j^V$ is  similar to the BBGKY hierarchy appearing in the study of many particles hamiltonian systems.  Notice that in such cases the underlying microscopic dynamics is deterministic (since it is given by the Newton equations associated with the Hamiltonian under consideration), while the  dynamics of the coalescing particles is probabilistic, given by the Marcus-Lushnikov process. Nevertheless,  our system of countably coupled equations may be treated along  similar lines and our result may be stated in the same language in terms of propagation of chaos. These arguments inspired from many particles hamiltonian systems
have already been used in the context of particle's coalescence in \cite{Lang} where propagation of chaos is proved for a coagulation equation with spatial dependence and constant kernel.

Let us recall here that for mean field models of classical particles with suitable two body potential, the propagation of chaos holds true  \cite{MR0475547}, \cite{MR541637}, \cite{MR740721} and may be seen as a law of large numbers.  The method of the proof  is based on the use of the so called empirical distribution and,  as explained in \cite{MR1956958}, it  relies upon the two following facts: the empirical distribution is a weak solution of the Vlasov equation; weak solutions to the Vlasov equation are continuous with respect to the initial datum in the topology of the weak convergence of the measures.  In that case no proof of this result based on PDE arguments is known (cf. for example the discussion in  Section 1.4  of \cite{MR2656357}). 

When the coalescing kernel is constant, as it is considered in this work, the proof of the main result is rather simple and clear, due to the fact that the suitable functional space, the set of non negative integrable functions over $\R^+$,  is easily seen to be globally preserved along the time evolution of the system. The stochastic coalescent for $K=1$ is known as Kingman's coalescent. The
construction goes back to Kingman \cite{MR671034} and it has been extensively studied. Some details and references may be found in \cite{MR1673235} and \cite{MR2253162}. For more general kernels the proofs may become more technical and involved and some other delicate questions may appear (cf. \cite{PE}).

The plan of the paper is the following. In Section \ref{The particle system} we briefly recall the description of the finite particle system and present an existence and uniqueness result for the system of equations satisfied by the mass distribution functions. In Section \ref{SMarginals} we introduce the correlation functions and deduce the system of equations that they satisfy. Two simple time asymptotic properties of that system are derived in Section \ref{finitehierarchy}. The rescaled system is introduced in Section \ref{rescaled} where the main theorem is proved.

\section{The particle system}
\label{The particle system}
\setcounter{equation}{0}
\setcounter{theo}{0}

We want to describe a finite system of particles contained in a finite volume $V$ whose number $N$ is  not fixed through time. To this end we consider the probability space defined as follows.\\
The sample space is:
\bear
&&\Omega=\bigcup_{N=1}^{\infty}\left (N, \mathscr{P}_N(\mathbb{R}^+)\right)\subset  \mathbb{N^*}\times  \mathscr P({\mathbb R}^+)  \label{S2E1}\\
&&\mathscr P_N(\mathbb{R}^+): \hbox{subsets of} \,\, \mathbb{R}^+\,\,\,\hbox{with cardinal}\,\,N.\label{S2E1bis}
\eear
It is the set of all the pairs $(N, \omega _N)$ where $N\in \mathbb{N^*}$ and $\omega _N$ is any finite subset of $N$ positive real numbers  $m_1, m_2, \cdots, m_N$.  For every time $t\ge 0$ the probability distribution of each state $(N, (m_1, \dots,  m_N))\in \Omega$ is $P_N(t, m_1, \dots,  m_N)/N!$ where $\left\{P_N(t)\right\}_{N\in  \mathbb{N}^*}$ is a sequence of non negative functions $P_N(t)=P_N(t, \cdot)$, each of them  defined  on $\mathscr P_N({\mathbb R}^+)$ and normalized according to:
\bear
\label{S2E2}
\sum_{N=1}^\infty \frac{1}{N!}\int_0^\infty dm_1\cdots \int_0^\infty dm_N P_N(t, m_1, \cdots, m_N)=1
\eear
The functions $P_N(t, m_1, \dots, m_N)$ are assumed to be symmetric with respect to any permutation of the indices  $1,\dots, N$ and no  restrictions are imposed on  the range of mass values other than $m_i>0$ for $i=1,\dots, N$. Each $P_N=P_N(t, m_1, \dots, m_N)$ is 
the mass distribution  function of the $N$-particle configuration $(m_1, \dots, m_N)$. For every $N\in \mathbb{N}^*$ the function:
\bear
\label{S2E4}
P(t, N)= \frac{1}{N!}\int_0^\infty dm_1\cdots \int_0^\infty dm_N P_N(t, m_1, \cdots, m_N)
\eear
is the probability that at time $t$ the system is constituted by $N$ particles and the normalization (\ref{S2E2}) is natural.
\\
\begin{rem}
The factor $1/N!$ in definition (\ref{S2E4}) is needed to compensate for counting all the $N!$ physically equivalent ways of arranging the masses $m_1,\dots, m_N$ in order of size.
\end{rem}

Let $\mathcal{A}_V(m,\mu)\geq 0$ be the rate (per unit time) at which a single particle of mass $m$ collides and coalesces with a particle of mass $\mu$ (given only that each is in the considered spatial region of volume $V$). 
We assume that:
\begin{eqnarray}\label{eq:A}
\mathcal{A}_V(m,\mu)=\mathcal{A}_V(\mu, m),
\end{eqnarray}
as it is reasonable from a physical point of view.

In a coalescing system  the number of particles will be varying along the time evolution (actually it will be decreasing) so what is meaningful is to consider  the time evolution  of the sequence of mass distribution functions $\{P_N(t)\}_{N\in \N^\ast}$. 
As explained in \cite{MR0223151}, \cite{1972JAtS...29.1496G}, \cite{Tanaka1994404} and  \cite{Lushnikov1978276} 
the evolution equations for the mass distribution functions are:
\begin{eqnarray}\label{eq:PN}
\pa_t P_N &=& \frac{1}{2}\sum_{\ell=1}^N \int_0^{m_\ell} d\mu\ \mathcal{A}_V(m_\ell - \mu, \mu) P_{N+1}(m_1, \dots, m_{\ell -1}, m_\ell - \mu, m_{\ell +1}, \dots, m_N, \mu,t)+\nonumber\\
&-& \frac{1}{2}\sum_{\ell\neq q }^N P_N(m_1, \dots, m_N,t)\mathcal{A}_V(m_\ell, m_q).\qquad N=1,2,\dots
\end{eqnarray}
The first term on the right hand side of (\ref{eq:PN}) is the so-called gain term and it describes the positive contribution due to the coagulation  of a particle of mass $\mu$, with $\mu\leq m_\ell$, with a particle of mass $m_\ell - \mu$ (in an $N+1$-particle configuration), giving rise to a particle of mass $m_\ell$ (in an $N$-particle configuration). 
The second term on the right hand side of (\ref{eq:PN}) is the so-called loss term and it describes the negative contribution due to the coagulation  of a particle of mass $m_\ell$ with a particle of mass $m_q$ (in an $N$-particle configuration), giving rise to a particle of mass $m_\ell+m_q$ (in an $N-1$-particle configuration). 
That set of equations completely  neglects the contributions that would be due either to many-boby collisions (e.g., three-body collisions passing from an $N+2$ configuration to an $N$ configuration or from an $N$ configuration to an $N-2$ configuration ), or to the occurrence of multiple binary collision (e.g., double binary collision again passing from an $N+2$ (or $N$) configuration to an $N$ (or $N-2$) configuration). 

An important feature of the system (\ref{eq:PN}) is that the total mass is preserved. More precisely, denoting by $M_N$ the total mass of the configuration $m_1,\dots, m_N$, i.e. $m_1+\dots + m_N=M_N$, then the only processes that have a non zero contribution in the time variation of the distribution $P_N(m_1,\dots, m_N, t)$ are those associated with configurations with the same total mass $M_N$. In fact, the gain term in (\ref{eq:PN}) takes into account $N+1$-particle configurations with total mass $m_1+\dots + m_\ell - \mu+\dots+m_N +\mu=M_N$ and the loss term in (\ref{eq:PN}) leads to $N-1$ particle configurations in which the total mass is $m_1+\dots+(m_\ell+m_q)+\dots+m_N=M_N$. 

The system of equations (\ref{eq:PN}) has been studied in  \cite{MR0223151}, \cite{1972JAtS...29.1496G}, \cite{Tanaka1994404} and  \cite{Lushnikov1978276}. For the sake of completeness we present in this Section an existence and uniqueness result that suits our purposes.

\subsection{The coalescence rate}
In general the coagulation kernel  $\mathcal{A}_V(m_1,m_2)$  at which two particles of masses $m_1$ and $m_2$ coalesce in a volume $V$,  has the following form:
\begin{equation}\label{scaleA0}
\mathcal{A}_V(m_1,m_2)= C_V\, \mathcal{A}(m_1,m_2).
\end{equation}
The function $\mathcal{A}(m_1,m_2)$ encodes only the dependence of the coagulation rate on the masses $m_1$ and $m_2$. The term $C_V$ contains the dependence of that rate with respect to the volume $V$ as well as that on other physical properties of the coalescence process under consideration. An example (considered for instance in  \cite{MR0223151}  and \cite{1972JAtS...29.1496G}) is the following:
\begin{equation}
\label{AMarcus}
C_V
=\frac{1}{V}E(|v_1 - v_2|)   \pi 
\end{equation}
where $E(|v_1 - v_2|)$ is 
the \emph{average relative velocity} of the two particles.
The volume dependence of $C_V$ takes into account the fact that the coagulation rate increases as the proportion of the volume occupied by the particles with respect to the total volume $V$ increases. That gives a dependence inversely proportional to $V$, i.e. like  $1/V$.  We consider in this paper the simplest possible case compatible with our purpose, that is to study the limit of the finite particle system
as $V\to +\infty$ at fixed density, namely:
\bear
\label{scaleA2}
\mathcal{A}_V(m_1,m_2)=\frac{1}{V}
\eear
Other kernels are considered in \cite{PE}.

\subsection{ The initial data}
We consider system  (\ref{eq:PN}) with  initial datum:

\begin{equation}
\label{PNindat0}
P_{N}^0(\mathbf{m}_{N})=\left\{
\begin{array}{l}
 (N_0)!\ f_0(m_1)\dots f_0(m_{N_0})\,,\,\,\hbox{if}\, \, N= N_0,\\ \\
0 \,\,\hbox{if}\, \, N\neq N_0,
\end{array}
\right.
\end{equation}
where the function $f_0$ is such that:
\begin{eqnarray}\label{indatSINGLENUMfin0}
f_0(m)\geq 0\ \ \ \hbox{a.e},\ \ \ \int_0^\infty dm\, f_0(m)=1.
\end{eqnarray}
By choosing the initial datum as in (\ref{PNindat0})-(\ref{indatSINGLENUMfin0}), and defining:
\begin{eqnarray}\label{PN0N00}
P^0(N)=\frac{1}{N!}\int_0^{\infty} d m_{1}\dots \int_0^{\infty} d m_{N}\  P^0_{N}(m_1, \dots, m_N),
\end{eqnarray}
for every $N\in\mathbb N^*$, we have
\begin{eqnarray}\label{PN01N00}
\sum_{N=1}^\infty \, P^0(N)=1
\end{eqnarray}
and condition (\ref{S2E2}) is satisfied at time $t=0$. With that choice of initial datum:
\bean
\label{S2E5}
P^0(N)=\delta(N-N_0),
\eean
i.e. at $t=0$ our system has exactly  $N_0$ particles.
\\
\subsection{Well-posedeness of the equation for $P_N(t)$ for the constant kernel}
For every $N \in\mathbb N^*$  we define the operator $G_N$,
mapping $N+1$-particle functions into $N$-particle functions, as follows:

\begin{eqnarray}\label{gainN}
G_N [Q_{N+1}](\mathbf{m}_{N})&=&\frac{1}{2V}\sum_{\ell=1}^N \int_0^{m_\ell} dm_{N +1}\ 
Q_{N+1}(m_1, \dots, m_{\ell -1}, m_\ell - m_{N +1}, m_{\ell +1}, \dots, m_N, m_{N +1}),\nonumber\\
&&
\end{eqnarray}
where $Q_{N+1}\in L^1((\R^+)^{N+1})$ and we denote:
\bear
\label{S2EvectorN}
\mathbf{m}_N:=(m_1, \dots, m_N).
\eear
We shall also use the following notation:
\bear
\label{S2EintN}
\int d\mathbf{t}_{N}:=\int_0^{t}dt_1\dots \int_0^{t_{N-1}}dt_{N}.
\eear
and denote $\mathcal H$ the set:
\bear
\mathcal H=\left\{\{f_N\}_{N\in\mathbb N^*}; \ f_k\in L^1((\R^+)^k),\,\,k\in \mathbb N^* \right\}
\eear

\begin{defin}
We say that a sequence $\{P_N(t) \}_{N\in \mathbb N^*}\in \mathcal H$ solves the system (\ref{eq:PN}) if, for every $t>0$ and every $N\in \mathbb N^*$, 
each term in 
(\ref{eq:PN}) belongs to $L^1((\R^+)^N)$ and the equality holds in $L^1((\R^+)^N)$.
\end{defin}
\begin{lem}
\label{S2Tlem}
For every $Q_{N+1}\in L^1((\R^+)^{N+1})$:
\bear
\left|\left|G_N[Q_{N+1}]\right|\right|_{L^1((\R^+)^{N})}\le \frac{N}{2V}\left|\left|Q_{N+1}\right|\right|_{L^1((\R^+)^{N+1})}. \label{S2Tlem1}
\eear
If $Q_{N+1}\in L^1((\R^+)^{N+1})$ and $Q_{N+1}\ge0$ then
\bear
\left|\left|G_N[Q_{N+1}]\right|\right|_{L^1((\R^+)^{N})}= \frac{N}{2V}\left|\left|Q_{N+1}\right|\right|_{L^1((\R^+)^{N+1})}.\label{S2Tlem2}
\eear
\end{lem}
\textbf{Proof of Lemma \ref{S2Tlem}} 
We start proving (\ref{S2Tlem1}). To this end we write:
\bear
&&\left|\left|G_N[Q_{N+1}]\right|\right|_{L^1((\R^+)^{N})}=
\int_0^\infty dm_1\cdots \int_0^{\infty}dm_N\left|G_N[Q_{N+1}] (m_1,\cdots, m_N)\right| \nonumber \\
&&\le  \frac{1}{2V}\sum_{\ell=1}^N\int_0^\infty\!\!\! dm_1\cdots \int_0^{\infty}\!\!\! dm_N\int_0^{m_{\ell}}dm_{N+1}
|Q_{N+1}(m_1, m_2, \cdots, m_{\ell}-m_{N+1},\cdots, m_{N+1})|  \label{S2lem1}\\
&&=
\frac{1}{2V}\sum_{\ell=1}^N\underbrace{\int_0^\infty \!\!\! dm_1\cdots \int_0^\infty \!\!\! dm_N}_{m_i\not = m_\ell}
\int_0^{\infty}\!\!\!dm_\ell\int_0^{m_{\ell}}\!\!\!dm_{N+1}
|Q_{N+1}(m_1, m_2, \cdots, m_{\ell}-m_{N+1},\cdots, m_{N+1})|\nonumber
\eear
where in the last step we have used that the integrals with respect to the variables $m_i$ for $i=1, \cdots, N$ are over independent domains and therefore can be performed in any order. The final and trivial step is to use Fubini's theorem in the integrals with respect to $m_\ell$ and $m_{N+1}$. That may be done since by hypothesis $Q_{N+1}\in L^1((\R^+)^{N+1})$ and therefore, for almost every $(m_1, m_2, \cdots, m_{\ell-1}, m_{\ell+1}, \cdots, m_N)$ 
$Q_{N+1}(m_1, m_2, \cdots, m_{\ell},\cdots, m_N, m_{N+1})\in L^1((\R^+)^2)$.  

If we also had $Q_{N+1}\ge0$ the inequality in (\ref{S2lem1}) would be an equality and we would have (\ref{S2Tlem2}).
\qed

We can state now the existence and uniqueness result of solutions for system (\ref{eq:PN}).
\begin{theo}
\label{S2T1}
Suppose that the $\mathcal A_V=V^{-1}$. Then,  the Cauchy problem for the system (\ref{eq:PN}) with initial data $\{P_N^0\}_{N\in {\mathbb N^*}}$ defined in (\ref{PNindat0})-(\ref{indatSINGLENUMfin0}) has a solution  $\{P_N (t)\}_{N\in \mathbb N ^*}\in \mathcal H$ for any $t$ and such that $P_N\in \mathbf C^\infty([0, +\infty), L^1((\R^+)^N))$ for every $N\in \mathbb N^*$. Such a solution is given by 
\begin{equation}
\label{S2TSolPN}
P_N (t, \mathbf{m}_N)=\left\{
\begin{array}{l}
\displaystyle{\int d\mathbf{t}_{N_0-N}}\, \left\{e^{-\frac{N(N-1)}{2V}\, (t-t_1)}\, G_N\right\}
\dots  \left\{e^{-\frac{(N_0-1)(N_0-2)}{2V}\, (t_{N_0 - N-1}-t_{N_0 - N})}G_{N_0-1}\right\}  \\  
\hskip 4cm e^{-\frac{(N_0)(N_0-1)}{2V}\, t_{N_0 - N}} P^0_{N_0}(\mathbf{m}_{N_0}),  \quad \hbox{for}\,\,1\le N< N_0\\ \\
e^{-\frac{(N_0)(N_0-1)}{2V}\, t} P^0_{N_0}(\mathbf{m}_{N_0}),  \quad \hbox{for}\,\,N= N_0\\ \\
 
 0, \,\,\,\,\hbox{for}\, \, \,\,N> N_0.
\end{array}
\right.
\end{equation}
Moreover, for every $t>0$, $P_N(t)\ge0$ and:
\begin{eqnarray}
&&\sum_{N=1}^\infty \, P(t, N)=1\label{PN02N0}\\
\text{with}\ &&P(t, N):=\frac{1}{N!}\int_0^\infty dm_1\cdots \int_0^\infty dm_N P_N(t, m_1, \cdots, m_N)\label{PN02N0bis}
\end{eqnarray}
That solution is unique in the set of sequences $\{\overline{P}_N(t)\}_{N\in \mathbb N^*}\in \mathcal H$ such that $\overline{P}_N\in \mathbf C^1([0, +\infty); L^1((\R^+)^{N}))$, $\overline{P}_N\ge 0$ for all $N\in \mathbb N^*$ and the function $h_{\overline{P}}(t):=\sum_{N=1}^\infty \overline{P}(t, N)$ satisfies
$h_{\overline{P}}\in \mathbf C^1([0, +\infty))$.
\end{theo}

\textbf{Proof of  Theorem \ref{S2T1}} By our hypothesis $P^0_{N_0}\in L^1((\R^+)^{N_0})$. By Lemma (\ref{S2Tlem}) we deduce that for every $1\le N\le N_0$:
\bean
G_N\, G_{N+1}
\dots G_{N_0-1} [P^0_{N_0}]\in L^1((\R^+)^{N}).
\eean
Then, for any $a$ and $\tau$, the function
\bean
\eta (t, \mathbf m_N)= e^{a (t-\tau)}G_N\, G_{N+1}
\dots G_{N_0-1} [P^0_{N_0}](\mathbf m_N)
\eean
satisfies:
\bean
\eta \in \mathbf C^{\infty}([0, +\infty),\,\,L^1((\R^+)^{N})).
\eean
It then follows that $P_N(t)$ given by (\ref{S2TSolPN}) is such that $P_N\in \mathbf C^\infty([0, +\infty), L^1((\R^+)^N))$. 

We check now that the above sequence $\{P_N(t)\}_{N\in \mathbb N^*}$ satisfies the system (\ref{eq:PN}) . The equations  for $N\ge N_0+1$ are trivially satisfied. Suppose now that $N\le N_0-1$.  Taking the time derivative of the left hand side of (\ref{S2TSolPN}) we obtain:
\bean
\frac{\partial P_N}{\partial t}(t, \mathbf m_N)&=&-\frac{N(N-1)}{2V}P_N(t, \mathbf m_N)+\\
&&\int_0^tdt_2\int_0^{t_2}dt_3 \dots \int_0^{t_{N_0 - N-1}} dt_{N_0 - N}\, G_N \left\{e^{-\frac{(N+1)N}{2V}\, (t-t_2)}G_{N_0-1} \right\}
\dots  \\
&&\hskip 2cm \dots\left\{e^{-\frac{(N_0-1)(N_0-2)}{2V}\, t_{N_0 -N-1}}G_{N_0-1} \right\} 
\ \ e^{-\frac{(N_0)(N_0-1)}{2V}\, t_{N_0 - N}} P^0_{N_0}(\mathbf{m}_{N_0})\\
&=&-\frac{N(N-1)}{2V}P_N(t, \mathbf m_N)+\\
&&G_N\left[\int_0^tdt_1\int_0^{t_1}dt_2 \dots \int_0^{t_{N_0 - N-2}} dt_{N_0 - N-1}\, \left\{e^{-\frac{(N+1)N}{2V}\, (t-t_1)}G_{N_0-1} \right\}
\dots  \right.\\
&&\left.\hskip 2cm\dots \left\{e^{-\frac{(N_0-1)(N_0-2)}{2V}\, t_{N_0 - N-2}}G_{N_0-1} \right\}
\ e^{-\frac{(N_0)(N_0-1)}{2V}\, t_{N_0 - N-1}} P^0_{N_0}(\mathbf{m}_{N_0})\right]
\eean
And, since
\bean
&&\int_0^tdt_1\int_0^{t_1}dt_2 \dots \int_0^{t_{N_0 - N-2}} dt_{N_0 - N-1}\, \left\{e^{-\frac{(N+1)N}{2V}\, (t-t_1)}G_{N_0-1} \right\}
\dots \\
&& \hskip 1.5cm\dots \left\{e^{-\frac{(N_0-1)(N_0-2)}{2V}\, t_{N_0 - N-2}}G_{N_0-1} \right\}
\  e^{-\frac{(N_0)(N_0-1)}{2V}\, t_{N_0 - N-1}} P^0_{N_0}(\mathbf{m}_{N_0})\equiv P_{N+1}(t, \mathbf m_{N+1})
\eean
 we obtain
\bean
\frac{\partial P_N}{\partial t}(t, \mathbf m_N)=-\frac{N(N-1)}{2V}P_N(t, \mathbf m_N)+G_N[P_{N+1}](t, \mathbf m_N)
\eean
that is nothing but the $N^{\hbox{th}}$ equation of system (\ref{eq:PN}). If $N=N_0$, we have:
\bean
\frac{\partial P_{N_0}}{\partial t}(t, \mathbf m_{N_0})&=&-\frac{N_0(N_0-1)}{2V}P_N(t, \mathbf m_{N_0})\\
&\equiv&-\frac{N_0(N_0-1)}{2V}P_N(t, \mathbf m_{N_0})+G_{N_0}[P_{N_0+1}](t, \mathbf m_{N_0})
\eean
since, by (\ref{S2TSolPN}), $P_{N_0+1}(t)\equiv 0$ for any $t$.  

We check now (\ref{PN02N0}). To this end we first notice that, by (\ref{S2TSolPN}) :
\begin{eqnarray*}
\sum_{N=1}^\infty \, P(t, N)&=&\sum_{N=1}^{N_0}\frac{1}{N!}\int_0^{\infty} d m_{1}\dots \int_0^{\infty} d m_{N}\  P_{N}(t, m_1, \dots, m_N).
\end{eqnarray*}
Therefore, using (\ref{eq:PN}) and (\ref{S2Tlem2}) (since $P_N\ge 0$):
\bean
\frac{d }{dt}\sum_{N=1}^\infty \, P(t, N)&=&-\sum_{N=1}^{N_0}\frac{N(N-1)}{N!\, 2V}||P_N(t)||_{L^1((\R^+)^N)}+
\sum_{N=1}^{N_0}\frac {1} {N!}||G_N[P_{N+1}]||_{L^1((\R^+)^N)}\\
&=&-\sum_{N=1}^{N_0}\frac{N(N-1)}{N!\, 2V}||P_N(t)||_{L^1((\R^+)^N)}
+\sum_{N=1}^{N_0-1}\frac{N}{N! 2V}||P_{N+1}(t)||_{L^1((\R^+)^{N+1})}\\
&=&-\sum_{N=1}^{N_0}\frac{N(N-1)}{N!\, 2V}||P_N(t)||_{L^1((\R^+)^N)}
+\sum_{M=2}^{N_0}\frac{M-1}{(M-1)! 2V}||P_{M}(t)||_{L^1((\R^+)^{M})}\\
&=&-\sum_{N=1}^{N_0}\frac{N(N-1)}{N!\, 2V}||P_N(t)||_{L^1((\R^+)^N)}
+\sum_{M=1}^{N_0}\frac{M(M-1)}{M! 2V}||P_{M}(t)||_{L^1((\R^+)^{M})}\equiv 0.
\eean
Identity (\ref{PN02N0}) follows from the fact that, by continuity and  (\ref{PN01N00}) one has:
\bean
\sum_{N=1}^{N_0} \,P(t, N)=\sum_{N=1}^{N_0} \, P^0(N)=1.
\eean

In order to prove the uniqueness, we first show that any solution $\{\overline{P}_N(t)\}_{N\in \mathbb N^*}$,  
such that  $\overline{P}_N(t)\ge 0$ 
and 
$h_{\overline{P}}(t)=\sum_{N=1}^\infty \overline{P} (t,N)$ 
verifies $h_{\overline{P}}\in C^1((0, +\infty))$, satisfies $\overline{P}_N(t) \equiv 0$ for all $t$ and $N\ge N_0+1$. To this end we write:
\bean
\frac{d}{dt}\sum_{N=N_0+1}^\infty \overline{P}(t, N)&=&-\sum_{N=N_0+1}^\infty \frac{N(N-1)}{N!2V}||\overline{P}_N(t)||_{L^1((\R^+)^N)}+\sum_{N=N_0+1}^\infty
||G_N[\overline{P}_{N+1}]||_{L^1((\R^+)^{N})}\\
&=&-\sum_{N=N_0+1}^\infty \frac{N(N-1)}{N!2V}||\overline{P}_N(t)||_{L^1((\R^+)^N)}+\sum_{N=N_0+1}^\infty
\frac{N}{N! 2V}||[\overline{P}_{N+1}]||_{L^1((\R^+)^{N+1})}\\
&=&-\sum_{N=N_0+1}^\infty \frac{N(N-1)}{N!2V}||\overline{P}_N(t)||_{L^1((\R^+)^N)}+\sum_{N=N_0+2}^\infty
\frac{N(N-1)}{N! 2V}||[\overline{P}_{N}]||_{L^1((\R^+)^{N})}\\
&=&-\frac{N_0(N_0+1)}{(N_0+1)!2V}||\overline{P}_{N_0+1}(t)||_{L^1((\R^+)^{N_0+1})}<0
\eean
We deduce that:
\bear
\label{S2EW1}
\sum_{N=N_0+1}^\infty \overline{P}(t, N)\le \sum_{N=N_0+1}^\infty \overline{P}(0, N)=\sum_{N=N_0+1}^\infty P^0( N).
\eear
Since, by hypothesis, the right hand side of  (\ref{S2EW1}) is zero and $\overline{P}_N(t)\ge 0$ we deduce that $\overline{P}_N(t)\equiv 0$ for every $N\ge N_0+1$ and $t\ge 0$. That leaves only a finite number of equations in the system  (\ref{eq:PN}), for $N=1, \cdots, N_0$. Moreover the equation for $N=N_0$  yields:
\bean
P_{N_0}(t, \mathbf m_{N_0})=P^0_{N_0}(\mathbf m_{N_0})e^{-\frac{N_0(N_0-1)}{2V} t}.
\eean
Then, the finite system decouples and may be explicitly solved for $N=1,\cdots, N_0-1$ to obtain expression (\ref{S2TSolPN}).
\qed 
\begin{rem}
One would expect that for every initial data $\{P_N^0\}_{N\in \mathbb N^*}\in \mathcal H$ such that $P_N^0\ge 0$ for every $N$ and $\sum P^0(N)=1$  there exists a solution  $\{P_N(t)\}_{N\in \mathbb N^*}$ such that, for every $N$:  $P_N(t)\ge 0$,  $P_N\in \mathbf C([0, +\infty); L^1((\R^+)^N))\cap \mathbf C^1((0, +\infty); L^1((\R^+)^N))$,
$h_P\in \mathbf C^1(0, +\infty)$ and $\sum_{N=1}^\infty P(t, N)=1$.  Since such a result is not necessary for our main purpose, the proof  of Theorem \ref{MAINTh}, we only prove  the simpler result in Theorem \ref{S2T1}.
\end{rem}

\section{Correlation Functions}
\label{SMarginals}
\setcounter{equation}{0}
\setcounter{theo}{0}
For any fixed $j\in \N^\ast$ , we define the $j$-particle correlation function $f_j(m_1, \dots, m_j,t)$ 
at time $t$ as
\begin{eqnarray}\label{eqMARGINALt=0}
f_j(m_1,\dots, m_j,t) = \sum_{N=j}^\infty \frac{1}{(N-j)!}\int_0^{\infty} d m_{j+1}\dots \int_0^{\infty} d m_{N}\  P_{N}(m_1, \dots, m_N,t).
\end{eqnarray}
Some well known general properties of such functions $f_j$ are the following. At any time $t$ the expected (or mean) number of particles $\overline {N}(t)$ defined as
\bean
\overline{N(t)}=\sum_{N=1}^\infty N P(t, N)
\eean
satisfies:
\bear
\label{S3E100}
\overline{N(t)}=\int_0^\infty f_1(m_1, t)dm_1.
\eear
The function $f_1$ is then the density function associated to the average number of particles. More generally one may define:
\bean
\overline{(N(t))(N(t)-1)\dots (N(t)-j+1)}=\sum_{N=1}^\infty N(N-1)\dots (N-j+1) P(t, N)
\eean
and one has
\bear
\label{S4DefMj}
\overline{(N(t))(N(t)-1)\dots (N(t)-j+1)}=||f_j||_{L^1((\R^+)^j)}
\eear
The functions $f_j$'s will be called  correlation functions since their definition is very similar to that of the classical correlation functions in statistical mechanics (cf. \cite{MR0289084}) and satisfy properties  (\ref{S3E100}) and (\ref{S4DefMj}). Notice nevertheless that they slightly differ from the correlation functions in statistical mechanics since in particular, there is no activity parameter (cf. \cite{MR0289084}).

As an immediate Corollary of Theorem \ref{S2T1} we have the following.
\begin{cor}
\label{S3T1}
Consider the solution $\{P_N(t)\}_{N\in \mathbb N^*}$ of the Cauchy problem (\ref{eq:PN})-(\ref{PNindat0})-(\ref{indatSINGLENUMfin0}), whose existence and uniqueness have been proved in Theorem \ref{S2T1}. Then, for every $j\in \N^*$ the functions $f_j (t)$ defined as
\bear
\label{S3T1E2}
f_j(m_1\, \cdots, m_j, t)=\sum_{N=j}^\infty \frac{1}{(N-j)!}\int_0^{\infty} d m_{j+1}\dots \int_0^{\infty} d m_{N}\  P_{N}(t, m_1, \dots, m_N)
\eear
are such that:
\bear
&&(i)\quad f_j(0)=\frac{(N_0)!}{(N_0-j)!}\ f_0^{\otimes j},\,\,\, \forall\,\,j\in \{1, \dots, N_0\}
; \quad f_j(0)\equiv 0, \,\,\,
\forall  j\ge N_0+1. \label{S3T1E0}\\
&&(ii)\quad f_j(t)\ge 0\,\,\, \forall \,\,j\ge 1\ and\ 
\forall \ t\ge 0; \quad f_j(t)\equiv 0\ 
\forall \ t\ge 0 \,\,\,\hbox{if}\,\,j\ge N_0+1. \label{S3T1E3}\\
&&(iii)\quad f_j\in \mathbf C^\infty ([0, +\infty); L^1((\R^+)^j)).\ \ \label{S3T1E4}\\
&&(iv)\quad ||f_j(t)||_{L^1((\R^+)^j)}\le 2^{j-1}N_0! \left(\frac{t}{V}\right)^{N_0-j}\sum_{N=0}^{N_0-j}\frac{2^N}{N!}t^{-N}V^N
\,\,\,\hbox{for}\,\,j=1,\cdots, N_0.\label{S3T1E6}
\eear
\end{cor}
\textbf{Proof of Corollary \ref{S3T1}} Only (\ref{S3T1E6}) needs an explanation. It follows from the fact that, by definition:
\begin{eqnarray}\label{eqL1margSPEC}
||f_j(t)||_{L^1((\R^+)^j)} &=&\sum_{N=j+1}^{N_0} \frac{1}{(N-j)!}|| P_{N}(t)||_{L^1((\R^+)^N)},
\end{eqnarray}
where we used that $P_N(t)\equiv 0$ for $N\geq N_0 +1$.
On the other hand, by Theorem  \ref{S2T1}, for every $N\le N_0$
\begin{eqnarray*}
&&|| P_{N}(t)||_{L^1((\R^+)^N)}=\int d\mathbf{t}_{N_0 - N}\int d\mathbf{m}_N e^{-\frac{N(N-1)}{2V} (t-t_1)} G_N
\dots G_{N_0-1}  e^{-\frac{(N_0)(N_0-1)}{2V}\, t_{N_0 - N}} P_{N_0}(\mathbf{m}_{N_0}, 0 ).
\end{eqnarray*}
Using (\ref{S2Tlem2}) in Lemma \ref{S2Tlem} we deduce:
\bear
\label{eq:PN12new}
|| P_{N}(t)||_{L^1((\R^+)^N)}&\leq& \frac{t^{N_0 - N}}{(N_0 - N)!}\, 2^{-(N_0 - N)}\ V^{-(N_0 - N)} \, N(N+1)\dots (N_0-1)\left\| P_{N_0}(0)\right\|_{L^1((\R^+)^{N_0})}=\nonumber\\
&=&\frac{t^{N_0 - N}}{(N_0 - N)!}\, 2^{-(N_0 - N)}\ V^{-(N_0 - N)} \, \frac{(N_0-1)!}{(N-1)!}\, \left\| P_{N_0}(0)\right\|_{L^1((\R^+)^{N_0})}.
\end{eqnarray}
Using the inequality
\bear
\label{S3E101}
\,\,\,\forall a\ge b\ge 1:\quad \frac{(a-1)!}{(a-b)!(b-1)!}\leq  2^{a-1},
\eear
we obtain from (\ref{eq:PN12new}):
\bear
\label{eq:PN12}
|| P_{N}(t)||_{L^1((\R^+)^N)}&\leq& \frac{t^{N_0 - N}}{(N_0 - N)!}\, 2^{-(N_0 - N)} \ V^{-(N_0 - N)}\, 2^{N_0-1}\, (N_0 - N)!\, \left\| P_{N_0}(0)\right\|_{L^1((\R^+)^{N_0})}\nonumber\\
&=&2^{N-1}\, t^{N_0 - N}\, V^{-(N_0 - N)}\ N_0!,
\end{eqnarray}
where we used assumptions (\ref{PNindat0})-(\ref{indatSINGLENUMfin0}) on the initial datum $P_{N_0}(0)$.  Estimate  (\ref{S3T1E6}) follows from (\ref{eqL1margSPEC}) and (\ref{eq:PN12}).\qed\\

\subsection{The system of equations for the functions $f_j$}
An immediate consequence of the above calculation and Corollary (\ref{S3T1}) is the following:\\
\begin{theo}
\label{S3T2}
Suppose that $\{P_N(t)\}_{N\in  \N^*}$ is the unique solution of  (\ref{eq:PN}) with initial data $\{P_N^0\}_{N\in  \N^*}$ defined in (\ref{PNindat0})-(\ref{indatSINGLENUMfin0}) and coagulation kernel $\mathcal A_V=V^{-1}$.
 Then the set of functions $\{f_j(t)\}_{j\in \N^*}$ defined in (\ref{S3T1E2}) solves the following problem:
 \begin{eqnarray}\label{BBGKYtris}
\pa_t f_j &=& \frac{1}{2V}\sum_{\ell=1}^j \int_0^{m_\ell} d\mu\ f_{j+1}(m_1, \dots, m_\ell - \mu, \dots, m_j, \mu,t)+\\
&-& \frac{1}{2V}\ \frac{j(j-1)}{2}\ f_j(m_1, \dots, m_j,t)-\frac{j}{V}\,  \int_0^{\infty} d \mu\,  f_{j+1}(m_1, \dots, m_j, \mu,t),
\quad j\in \N^*,\nonumber
\end{eqnarray}
Moreover the sequence   $\{f_j(t)\}_{j\in \N^*}$ is the unique solution  in $\mathbf C([0, +\infty); L^1((\R^+)^j))\cap \mathbf C^1((0, +\infty); L^1((\R^+)^j))$  of  (\ref{BBGKYtris}) with initial data given   (\ref{S3T1E0}).
\end{theo}
\textbf{Proof of Theorem \ref{S3T2}} 
A straightforward computation from (\ref{eq:PN}) using the definition (\ref{S3T1E2}) shows that for any $j\le N_0$ the function $f_j$ satisfies the equation :
\begin{eqnarray}\label{BBGKY}
\pa_t f_j (\mathbf m_j, t)&=& \frac{1}{2V}\sum_{N=j}^{N_0} \frac{1}{(N-j)!}\sum_{\ell=1}^N \int d \mathbf{m}_{N, j}\int_0^{m_\ell} dm_{N+1}\nonumber \\
&&\hskip 6cm   P_{N+1}(m_1, \dots, m_\ell - m_{N+1}, \dots, m_{N+1},t)+\nonumber\\
&-& \sum_{N=j}^{N_0} \frac{1}{(N-j)!}\frac{N(N-1)}{2V}\int d \mathbf{m}_{N, j}\,   P_N(m_1, \dots, m_N,t),
\end{eqnarray}
where from now on we use the notation:
\begin{equation}\label{notation}
\mathbf{m}_{N, j}:= (m_{j+1},\dots,m_N)\ \ \ \text{and}\ \ \ \int d \mathbf{m}_{N, j}:=\int_0^{\infty} d m_{j+1}\dots \int_0^{\infty} d m_{N}.
\end{equation}
The first term in the right hand side of (\ref{BBGKY}), or gain term,  gives the following contributions:
\begin{eqnarray}\label{BBGKYgain}
&& \frac{1}{2V}\sum_{N=j}^{N_0} \frac{1}{(N-j)!}\sum_{\ell=1}^N \int d \mathbf{m}_{N, j} \int_0^{m_\ell} dm_{N+1}\  P_{N+1}(m_1, \dots, m_\ell - m_{N+1}, \dots,m_{N+1},t)\nonumber\\
&&=\frac{1}{2V}\sum_{\ell=1}^j \int_0^{m_\ell} dm_{N+1}\, f_{j+1}(m_1, \dots, m_\ell - m_{N+1}, \dots, m_j, m_{N+1},t)+\nonumber\\
&&+ \ \frac{1}{2V}\sum_{N=j}^{N_0} \frac{N-j}{(N-j)!} \int d \mathbf{m}_{N, j+1}\int_0^{\infty} d m_{j+1}\int_0^{m_{j+1}} dm_{N+1}\nonumber\\
&&\qquad\qquad \qquad\qquad\qquad\qquad P_{N+1}(\mathbf{m}_{j}, m_{j+1} -m_{N+1} , \dots,m_{N+1},t),
\end{eqnarray}
where we have split the sum with respect to $\ell$ in two parts, $1\le \ell\le j$ and $j+1\le \ell \le N$, and used the symmetry of $P_N$ with respect to any permutation of the indeces (that is preserved by the dynamics), to write:
\begin{eqnarray*}\label{BBGKYgain10bis}
&& \sum_{\ell=j+1}^N \int_0^{m_\ell} dm_{N+1}\ P_{N+1}(m_1, \dots, m_\ell - m_{N+1}, \dots, m_{N+1},t),\nonumber\\
&&=(N-j) \int_0^{m_{j+1}} dm_{N+1}\ P_{N+1}(m_1, \dots, m_{j+1} - m_{N+1},\dots, m_{N+1},t).
\end{eqnarray*}
Using Fubini's theorem and definition (\ref{S3T1E2}) we obtain
\begin{eqnarray}\label{BBGKYgain1}
&&\frac{1}{2V}\sum_{N=j}^{N_0} \frac{N-j}{(N-j)!} \int d \mathbf{m}_{N, j+1}\int_0^{\infty} \hskip -0.2cm d m_{j+1}
\int_0^{m_{j+1}}\hskip -0.6cm dm_{N+1} P_{N+1}(\mathbf{m}_{j}, m_{j+1} -m_{N+1} , \dots,m_{N+1},t)\nonumber \\
&&= \ \frac{1}{2V} \int_0^{\infty} d m_{N+1} \int_0^{\infty} dm_{j+1}\  f_{j+2}(m_1, \dots,m_{j+1}, m_{N+1},t).
\end{eqnarray}
On the other hand, by simple algebraic manipulations,  the  second  term in the right hand side of (\ref{BBGKY}), or loss term,  gives:
\begin{eqnarray}\label{BBGKYlost}
&&\sum_{N=j}^{N_0} \frac{1}{(N-j)!}\frac{N(N-1)}{2V} \int_0^{\infty} d m_{j+1}\dots \int_0^{\infty} d m_{N} P_N(m_1, \dots, m_N,t) \nonumber\\
&&= \ \frac{j(j-1)}{2V}  \, f_j(m_1, \dots, m_j,t)+\frac{j}{V} \int_0^{\infty} d m_{j+1}\,  f_{j+1}(m_1, \dots, m_{j+1},t)+\nonumber\\
&&   +\ \frac{1}{2V} \int_0^{\infty} d m_{j+1}\int_0^{\infty} d m_{j+2} \, f_{j+2} (m_1, \dots, m_{j+2},t),
\end{eqnarray}
By (\ref{BBGKY}), (\ref{BBGKYgain}), (\ref{BBGKYgain1}) and (\ref{BBGKYlost}),  (\ref{BBGKYtris}) follows.

Only  uniqueness remains to be proved. To this end we notice that
since $P_N(t)\equiv 0$ for every $N>N_0$, the same is true for the sequence
$\{f_j(t)\}_{j\in \N^*}$, namely, $f_j(t)\equiv 0$ for $j>N_0$. Therefore the system (\ref{BBGKYtris}) only contains a finite number of non trivial equations and, since the equation for $j=N_0$ only involves $f_{N_0}(t)$, the system decouples.  We conclude using  the same argument used in Theorem \ref{S2T1} to prove uniqueness for the mass distribution functions $\{P_N(t)\}_{N\in \N^*}$. \qed

\begin{rem}
\label{S3R300}
If the initial data $f_0$  satisfies  not only (\ref{indatSINGLENUMfin0}) but also:
\bean
m_0=\int _0^\infty mf_0(m)dm<+\infty
\eean 
then, since, as we have seen above,  $f_1(t)$ is the density function associated to the average number of particles at time $t$, the quantity:
\bear
\label{S3R300E1}
\overline{M(t)}=\int _0^\infty m\, f_1(m, t)\, dm
\eear
represents the average mass in the system at time $t$. 
If we integrate the equation for $j=1$ in system  (\ref{BBGKYtris}), a simple application of Fubini's theorem yields
 \bear
\label{S3R30E2}
\frac {d} {dt}\int _0^\infty m\, f_1(m, t)\, dm=0
\eear
and then, using (\ref{indatSINGLENUMfin0}) and (\ref{S3T1E0})
 \bear
\label{S3R30E3}
\overline{M(t)}=\overline{M(0)}=m_0\, N_0\quad\forall t>0.
\eear
This  mass conservation property is a well known feature of the coagulation equation with constant kernel. 
\end{rem}

\section{The solution of the finite system for $\{f_j(t)\}_{j=1}^{N_0}$}
\label{finitehierarchy}
\setcounter{equation}{0}
\setcounter{theo}{0}
In this Section we study the behavior of the sequence $\{f_j(t)\}_{j=1}^{N_0}$  as $t\to +\infty$.
\begin{prop}
\label{S4Tprop}
Let $\{f_j(t)\}_{j=1}^{N_0}$ be  the solution of   (\ref{BBGKYtris}) with initial datum (\ref{S3T1E0}) given by Corollary \ref{S3T1}. Then
\bean
\lim_{t\to +\infty}\overline{N(t)}=1\,\,\,\hbox{and}\,\,\,\,\,
\lim_{t\to +\infty}var (N)(t)=0,
\eean
where $var (N)(t)= \overline{(N(t))^2} - \overline{N(t)}^2$ is the variance of the distribution on particle numbers $P(t,N)$.
\end{prop}
The proof of Proposition \ref{S4Tprop} follows from the  three following auxiliary results.
\begin{lem} 
\label{S4T1}
For all $t>0$ and $j$: 
\bean
\frac{d}{dt} ||f_j(t)||_{L^1((\R^+)^j)} = -\frac{j}{2V}||f_{j+1}(t)||_{L^1((\R^+)^{j+1})}-\frac{j(j-1)}{2\, V}||f_j(t)||_{L^1((\R^+)^j)}\leq 0.
\eean
\end{lem}
\textbf{Proof of Lemma \ref{S4T1}} Since, by definition, $f_j(t)\geq 0$ for any $j$ and $t$,  (the) Lemma \ref{S4T1} follows by a simple integration of (\ref{BBGKYtris}) with respect to $m_1\dots m_j$.
\qed
\vspace{0.5cm}

From  (\ref{S3E100}) and Lemma \ref{S4T1} we deduce
\begin{cor}  
\label{S4T2}
(i) For all $t>0$:
$$\frac{d}{dt}\,\overline{N(t)}\le 0$$
(ii) There exists $N_\infty\in [1, N_0]$ such that $\overline{N(t)}\to N_\infty$ as $t\to +\infty$.\\
\end{cor}
\textbf{Proof of Corollary \ref{S4T2}} Only  the fact $N_\infty\ge1$ needs perhaps to be explained. From $(i)$ we deduce the existence of $N_\infty\in [0, N_0]$, the unique limit point of $\overline {N(t)}$ as $t\to +\infty$.  On the other hand,
\bean
\overline {N(t)}=\sum_{N=1}^{N_0}N P(t, N)\ge  \sum_{N=1}^{N_0}P(t, N)=1
\eean
from where $N_\infty\ge 1$. 
\qed
\vspace{0.5cm}\\
We also deduce:
\begin{cor}
\label{S4T3}
For all $t>0$ and every $j\ge 1$:
\begin{eqnarray}\label{l1VfinjGENdecayNUM}
\overline{(N(t))(N(t)-1)\dots (N(t)-j+1)}\leq e^{-\frac{j(j-1)}{2\, V}\, t}\ N_0^j,
\end{eqnarray}
\end{cor}
\textbf{Proof of Corollary  \ref{S4T3}} By Lemma \ref{S4T1}:
\begin
{eqnarray}\label{l1VfinjGEN}
\frac{d}{dt} ||f_j(t)||_{L^1((\R^+)^j)} &= &-\frac{j}{2V}||f_{j+1}(t)||_{L^1((\R^+)^{j+1})}-\frac{j(j-1)}{2\, V}||f_j(t)||_{L^1((\R^+)^j)}\leq \nonumber\\
&\leq& \frac{j(j-1)}{2\, V}||f_j(t)||_{L^1((\R^+)^j)},
\end{eqnarray}
that implies, by (\ref{S3T1E0})
\begin{eqnarray}\label{l1VfinjGENdecay}
||f_j(t)||_{L^1((\R^+)^j)} &\leq &e^{-\frac{j(j-1)}{2\, V}\, t}\  \left\| f_j(0)\right\|_{L^1((\R^+)^j)}
\ \leq e^{-\frac{j(j-1)}{2\, V}\, t}\ N_0^j,
\end{eqnarray}
for any time $t$. Since, by definition, $\overline{(N(t))(N(t)-1)\dots (N(t)-j+1)}=||f_j(t)||_{L^1((\R^+)^j)} $,  \hfill \break this  concludes the proof.
\qed
\vspace{0.5cm}

\noindent
\textbf{Proof of Proposition \ref{S4Tprop}} By Corollary \ref{S4T3} for $j=2$:
\begin{eqnarray}\label{l1VfinDECAYj=2}
||f_2(t)||_{L^1((\R^+)^2)} =\overline{(N(t))(N(t)-1)}\leq e^{-\frac{t}{ V}}\ N_0^2\to 0,\ \ \ \text{as}\ \ t\to+\infty
\end{eqnarray}
and then
\begin{eqnarray}\label{l1VfinDECAYj=2BIS}
\lim_{t\to +\infty}\overline{(N(t))^2}=\lim_{t\to +\infty} \overline{N(t)}=N_\infty.
\end{eqnarray}
Therefore, the  
variance of the distribution on particle numbers for large times is given by:
\begin{eqnarray}\label{NUMofPART6reread2II}
&&\lim_{t\to +\infty}var(N)(t)=\lim_{t\to +\infty}(\overline{(N(t))^2}-\overline{N(t)}^2)=N_\infty-N_\infty^2.
\end{eqnarray}
Since, by definition, the variance is non-negative
\begin{eqnarray}\label{variance}
N_\infty-N_\infty^2\geq 0 \ \Rightarrow\ N_\infty(1-N_\infty)\geq 0 \ \Rightarrow\ 0\leq N_\infty\leq 1.
\end{eqnarray}
On the other hand by Corollary \ref{S4T2} we know that  $N_\infty \ge 1$, thus it has to be $N_\infty=1$
and therefore
\begin{eqnarray}\label{varianceII}
\lim_{t\to +\infty}var(N)(t)=0.
\end{eqnarray}
\qed

\begin{rem}
In other words, the distribution on particle numbers is going to be a delta distribution, as it was at time $t=0$, but now centered at  $N_\infty=1$. As expected, due to the coalescence dynamics, the system is going to be constituted by only one large particle of mass $N_0\, m_0$ .
\end{rem}

\section{BBGKY hierarchy of the rescaled correlation functions}

\label{rescaled}

\setcounter{equation}{0}
\setcounter{theo}{0}

Our purpose in this Section is to consider the limit where the volume $V$ and the initial number of particles 
$N_0$ go to infinity in such a way that 
\bear
\label{S5thermod}
\lim_{V, \,N_0 \to +\infty}\frac{N_0}{V}=\rho_0 \in (0, +\infty).
\eear
To this end let us define the rescaled correlation functions $\{f_j^V(t)\}_{j=1}^{N_0}$ as
\begin{eqnarray}\label{scalingK}
f_j^V(m_1,\dots, m_j,t)
:=  \frac{f_j(m_1,\dots, m_j,t)}{V^j},\ \ \ \ \ \  \ j=1,\dots, N_0.
\end{eqnarray}
Since, as we have seen in Section  \ref{SMarginals}, the function $f_1$ is the number density function, the rescaled function $f_1^V$ is the density function associated to the concentration of particles (number of particles per unit volume).
 
By (\ref{BBGKYtris})  the functions $f_j^V$ satisfy the following set of $N_0$ equations:
\begin{eqnarray}\label{BBGKY6}
\pa_t f_j^V &=& \frac{1}{2}\sum_{\ell=1}^j\int_0^{m_\ell} d\mu\ f_{j+1}^V(m_1, \dots, m_\ell - \mu, \dots, m_j, \mu,t)+\nonumber\\
&-& \frac{j(j-1)}{2V}\ f_j^V(m_1, \dots, m_j,t)-j\  \int_0^{\infty} d \mu\,  f_{j+1}^V(m_1, \dots, m_j, \mu,t),
\end{eqnarray}
for $j=1,2,\dots, N_0$.  We will refer to the family of equations (\ref{BBGKY6}) as BBGKY hierarchy 
by  analogy with the system arising  in the framework of many particles hamiltonian systems. 

By (\ref{S3T1E0}) and
 (\ref{scalingK})  the rescaled densities at time $t=0$ are
\begin{eqnarray}\label{scaling2}
f_j^V(m_1,\dots, m_j,0) =\frac{(N_0)!}{(N_0-j)!}\frac{1}{V^j}\ f_0^{\otimes j}(m_1,\dots, m_j),\ \ \ j=1,2,\dots, N_0
\end{eqnarray}
where the function $f_0$ has been defined in (\ref{indatSINGLENUMfin0}).
By (\ref{scaling2}), for every $j\ge 1$:
\bear
\label{scaling6}
\lim_{\substack{V,\,N_0 \to +\infty\\ \frac{N_0}{V}\to \rho_0}} ||f_j^V(0)-\rho_0^j f_0^{\otimes j}||_{L^1((\R^+)^j)}=0.
\eear

In order to state our main result  we first recall  that, for all non negative  initial data in $ L^1(\R^+)$,  the Cauchy problem for the coagulation equation (\ref{S1ECoagEqu}) with kernel equal to one
 has a unique non negative solution in $ \mathbf C([0, +\infty); L^1(\R^+))$  (cf.\cite{0944.60082}, Theorem 2.1).

Our main result is then the following:
\begin{theo}
\label{MainTh}
Let $\{f_j (t)\}_{j=1}^{N_0}$ be the solution of system  (\ref{BBGKYtris}) with initial data defined in  (\ref{S3T1E0}
). Then, if $\{f_j ^V(t)\}_{j=1}^{N_0}$ is the sequence of rescaled densities defined by (\ref{scalingK}):
\bear
\label{MainThE1}
\forall \ \ j\ge 1:\qquad \lim_{\substack{V,\, N_0\to +\infty\\ \frac{N_0}{V}\to \rho_0}} ||f_j^V(t)-
f(t)^{\otimes j}||_{L^1((\R^+)^j)}=0
\eear
where $f$ is the unique solution in $\mathbf C([0, +\infty); L^1(\R^+))$ of the coagulation equation (\ref{S1ECoagEqu}) with kernel $K=1$ and initial datum $\rho_0 f_0$, $f_0$  given by (\ref{indatSINGLENUMfin0}).
\end{theo}

The proof of Theorem \ref{MainTh} is done in two steps. One is to prove that  the sequence $\{f^V_j (t)\}_{j=1}^{N_0}$ converges to a sequence of functions $\{f_j^\infty (t)\}_{j\in \N^*}$ that satisfy an infinite set of equations. The second is to prove that for all  $j\ge 1$ and $t\geq 0$, $f_j^\infty(t)= f(t)^{\otimes j}$.  The first uses the explicit expression of the functions $f^V_j$ as a finite sum and a kind of dominated convergence. The second is done proving the uniqueness of solutions of the new infinite system of equations.  We actually start by proving this uniqueness result.

For the sake of notation  let us introduce operators $W_j$ defined as follows. Given a function $\varphi\in L^1((\R^+)^{j+1})$ we call $W_j$ the operator such that $W_j[\varphi]$ is:
\bear
\label{S6Wj}
W_j[\varphi](\mathbf m_j):= VG_j[\varphi](\mathbf m_j)-j\int_0^\infty d\mu\  \varphi (\mathbf m_j, \mu),
\eear
where $G_j$ is defined by  in (\ref{gainN}). Therefore,
using Lemma (\ref{S2Tlem}) it is easily seen that $W_j$ is a linear and continuous operator from $L^1((\R^+)^{j+1})$ to $L^1((\R^+)^j)$ whose norm satisfies:
\bear
\label{S6Wjnorma}
||W_j||\le \frac{3j}{2}.
\eear
\begin{lem}
\label{S6Tsmol}
Let $f\in \mathbf C([0, +\infty); L^1(\R^+))$ be the unique non negative solution of the coagulation equation
\begin{eqnarray}\label{Smoluchowski}
\pa_t f(m,t) &=& \frac{1}{2}\int_0^{m} d\mu\  f(m-\mu,t)f(\mu,t)- f(m,t) \int_0^{\infty} d \mu\,  f( \mu,t)
\end{eqnarray}
with initial datum $f(0, m)=\rho_0f_0(m)$, $f_0$ given by (\ref{indatSINGLENUMfin0}).  Then, the sequence of functions defined as
\bear
f_j^\infty(t, \mathbf m_j):=
f(t)^{\otimes j}(\mathbf m_j),\quad j\in \mathbb N^*\label{producto}
\eear
is the unique non negative solution of  the system:
\begin{eqnarray}\label{BBGKYjS}
\frac{\partial f_j^\infty}{\partial t}(\mathbf m_j,t) &=& \frac{1}{2}\sum_{\ell=1}^j\int_0^{m_\ell} d\mu\  f_{j+1}^\infty(m_1, \dots, m_\ell-\mu,\dots, m_j, \mu,t)+\nonumber\\
&-&  j\int_0^{\infty} d \mu\,  f_{j+1}^\infty(m_1,\dots,m_j, \mu,t), \ \ \text{for}\ \ j=1,2,\dots
\end{eqnarray}
 with initial data $f_j^\infty(0)=\rho_0^j f_0^{\otimes j}$ such that 
$f_j^\infty\in \mathbf C([0, +\infty); L^1(\R^+)^j)$.  For all $t>0$, the sequence $\{f_j^\infty(t)\}_{j\in \N^\ast}$ satisfies:
\bear
\label{eqSTIMAl1BISsmol}
\forall\  j\ge 1
:\quad ||f_j^\infty(t)||_{L^1((\R^+)^j)} \leq  \rho_0^j.
\eear
\end{lem}
\textbf{Proof of Lemma \ref{S6Tsmol}} A straightforward calculation shows that the sequence $\{f_j^\infty(t)\}_{j\in \mathbb N^*}$ defined in (\ref{producto}) is indeed a solution of system (\ref{BBGKYjS}) with initial data $\{\rho_0^j f_0^{\otimes j}(\mathbf m_j)\}_{j\in \mathbb N^*}$.

On the other hand, since $f\in \mathbf C([0, +\infty); L^1(\R^+))$ satisfies (\ref{Smoluchowski}) we have $f\in \mathbf C^1((0, +\infty); L^1(\R^+))$ and then  $f_{j}^\infty \in \mathbf C([0, +\infty); L^1(\R^+)^j)\cap  \mathbf C^1((0,  +\infty); L^1(\R^+)^j)$. Due to the fact that $\{f_j^\infty(t)\}_{j\in \mathbb N^*}$ satisfies  (\ref{BBGKYjS}), we obtain, after integration of the $j-$th equation over $(\R^+)^j$:
\begin{eqnarray*}\label{BBGKYjA=1l1Smol}
\frac{\partial }{\partial t} ||f_{j}^\infty(t)||_{L^1((\R^+)^j)} = -\frac{j}{2}||f_{j+1}^\infty(t)||_{L^1((\R^+)^{j+1})}\leq 0.
\end{eqnarray*}
and then, for all $t>0$:
\begin{eqnarray}
\label{eqSTIMAl1BISsmolPROOF}
 ||f_{j}^\infty(t)||_{L^1((\R^+)^j)}\le  ||f_{j}^\infty(0)||_{L^1((\R^+)^j)}= \rho_0^j.
\end{eqnarray}
That proves (\ref{eqSTIMAl1BISsmol}).

In order to prove uniqueness  
notice that the system (\ref{BBGKYjS}) may be written using operator $W_j$ as
\bean
\frac {\partial f^\infty_j} {\partial t}=W_j[f^\infty_j]
\eean
from where
\begin{eqnarray}\label{Duhamel1WS}
f_j^{\infty}(t)= f_j^{\infty}(0)+\int_0^t dt_1\, W_j \ f_{j+1}^{\infty}(t_1)
\end{eqnarray}
Using this same formula for $f^\infty_{j+1}$, we obtain after $M$ iterations:
\begin{eqnarray}\label{Duhamel2anteS}
f_j^{\infty}(t)&=& 
\sum_{n\geq 0}^M\ 
\frac{t^n}{n!}\ 
W_{j}\ W_{j+1}\ \dots W_{j+n-1}\  f_{j+n}^{\infty}(0)+\nonumber\\
&+& \int_0^{t}dt_1\dots\int_0^{t_{M}}dt_{M+1}\, W_{j}\dots 
W_{j+M}\ f_{j+M+1}^{\infty}(
\mathbf{m}_{j+M+1} ;t_{M+1}).
\end{eqnarray}

Let us assume now that there exist two different non-negative  solutions, $\{h_{j, k}(t)\}_{j\in \mathbb N^*}$, $k=1, 2$, of (\ref{BBGKYjS}) such that
$h_{j, k}\in \mathbf C([0, +\infty); L^1(\R^+)^j)$ with the same factorized initial datum $\{\rho_0^j\ (f_0)^{\otimes j}\}_{j\in \N^\ast}$. 
Then by (\ref{Duhamel2anteS}) and  (\ref{S6Wjnorma}) we deduce that for every $M>0$:
\begin{eqnarray}\label{Duhamel2anteUUS}
&&||h_{j,1}(t)-h_{j,2}(t)||_{L^1((\R^+)^j)}\leq \frac{t^{M+1}}{(M+1)!}\times\nonumber\\
&&\quad \times ||W_{j}\dots 
W_{j+M}\ \left(h_{j+M+1,\, 1}(
t_{M+1})-h_{j+M+1,\, 2}(t_{M+1})\right)||_{{L^1((\R^+)^{j+M+1})}} \nonumber\\
&&\le
\frac{t^{M+1}}{(M+1)!}\left( \frac{3}{2}\right)^{M+1} \frac{(j+M)!}{(j-1)!} \left (||h_{j+M+1,\, 1}(
t_{M+1})||_{{L^1((\R^+)^{j+M+1})}}\right.+ \nonumber\\
&&\hskip 8cm \left. +  || h_{j+M+1,\, 2}(t_{M+1})||_{{L^1((\R^+)^{j+M+1})}}\right),
\eear
By our assumptions on $\{h_{j, k}(t)\}_{j\in \mathbb N^*}$, $k=1, 2$, they both satisfy  (\ref{eqSTIMAl1BISsmol}).  Thus, plugging (\ref{eqSTIMAl1BISsmol}) into (\ref{Duhamel2anteUUS}) we deduce:
\bear
||h_{j,1}(t)-h_{j,2}(t)||_{L^1((\R^+)^j)}\leq 2 \rho_0^{j+M+1}\frac{t^{M+1}}{(M+1)!}\left( \frac{3}{2}\right)^{M+1} \frac{(j+M)!}{(j-1)!}
\eear
Using (\ref{S3E101}) we obtain
\bear
||h_{j,1}(t)-h_{j,2}(t)||_{L^1((\R^+)^j)}\leq 2^j \, \rho_0^j\, (3\rho_0 t)^{M+1}
\eear
and then, for every  $t< 1/(3 \rho_0)$:
\bear
\label{S6Uniqloc}
||h_{j,1}(t)-h_{j,2}(t)||_{L^1((\R^+)^j)}\leq \lim_{M\to +\infty}2^j \rho_0^j\,(3\rho_0 t)^{M+1}=0
\eear
and the uniqueness  follows but only for $t\in [0, 1/3\rho_0)$. This argument may be now iterated as follows. Suppose that  (\ref{S6Uniqloc}) holds for $t\in [0, T)$ for some $T>0$. Consider now  (\ref{Duhamel2anteUUS}) for $t\in [T, T+1/3\rho_0)$ (since it holds for all $t>0$).  Since (\ref{eqSTIMAl1BISsmol}) holds also true for all $t>0$, it follows as before that (\ref{S6Uniqloc}) also holds for $t\in [T, T+1/3 \rho_0)$. Therefore, global uniqueness is proven.
This ends the proof of Lemma  \ref{S6Tsmol}. \qed
\begin{lem}
\label{S6T1}
Under the same hypothesis than in Theorem \ref{MainTh} for all  $T>0$ and all $j\ge 1$: 
\bear
 \lim_{\substack{V,\, N_0 \to +\infty\\ \frac{N_0}{V}\to \rho_0}}\sup_{0\le t < T} ||f_j^V(t)-g_j(t)||_{L^1((\R^+)^j)}=0.\label{S6T1E2}
\eear
where 
\bear
\label{S6T1E100}
&&g_j(t, \mathbf m_j)=\sum_{n=0}^\infty \rho_0^{j+n}\frac{t^n}{n!}W_j\circ W_{j+1}\cdots \circ W_{j+n-1}[f_0^{\otimes {j+n}}](\mathbf m_j )
 \eear
 is such that $g_j \in \mathbf C([0, +\infty); L^1(\R^+)^j)\cap  \mathbf C^1((0,  \infty); L^1(\R^+)^j)$ and $g_j\ge 0$. Moreover the sequence $\{g_j\}_{j\in \N^*}$ satisfies system (\ref{BBGKYjS}) for $t\in (0, \infty)$.
\end{lem}

\textbf{Proof of Lemma \ref{S6T1}}
Let us prove first that the series in (\ref{S6T1E100}) defines a function $g_j \in \mathbf C([0, \tau_1); L^1(\R^+)^j)\cap  \mathbf C^1((0,  \tau_1); L^1(\R^+)^j)$ for some $\tau_1>0$.  To this end we first deduce the following estimate from (\ref{S6Wjnorma})  and the  the explicit expression (\ref{scaling2})  of the initial datum:
\bear
\label{S6EN1}
&&\left\|W_{j}\circ \cdots \circ W_{j+n-1}[f_0^{\otimes {j+n}}]\right\|_{L^1((\R^+)^{j})} \le \left(\frac{3}{2}\right)^n  j(j+1)\dots (j+n-1)\left\|f_0\right\|^{j+n}_{L^1(\R^+)}
\nonumber\\
&&=\left(\frac{3}{2}\right)^n  j(j+1)\dots (j+n-1).
\eear
from where,
using again (\ref{S3E101})
we deduce , for all $j\ge 1$, $n\ge 0$   and $t>0$:
\bear
&&\rho_0^{j+n}\, \frac{t^n}{n!}\left\|W_{j}\circ \cdots \circ W_{j+n-1}[f_0^{\otimes {j+n}}]\right\|_{L^1((\R^+)^{j})}\le 
2^{j-1}\,  \rho_0^j\, (3\rho_0 t)^n. \label{S6EM1}
\eear
It then follows that the series in (\ref{S6T1E100}) defines a continuous function $g_j\in \mathbf C([0, \frac{1}{3\rho_0}); L^1((\R^+)^j))$. A similar argument shows that $g_j\in \mathbf C^1((0, \frac{1}{3\rho_0}); L^1((\R^+)^j))$ and that
\bean
\frac {\partial g_j(t) } {\partial t}=\sum_{n=1}^\infty \rho_0^{j+n}\frac{t^{n-1}}{(n-1)!}W_j\circ W_{j+1}\cdots \circ W_{j+n-1}[f_0^{\otimes {j+n}}](\mathbf m_j ).
\eean
By construction the sequence $\{f_j^V(t)\}_{j=1}^{N_0}$ satisfies, for every $t>0$:
\begin{eqnarray}
\label{Duhamel1}
f_j^V(t, \mathbf m_j) &=&e^{-\frac{j(j-1)}{2\, V}t}f_j^V(\mathbf m_j;0)-\int_0^t dt_1\, e^{-\frac{j(j-1)}{2\, V}(t-t_1)} j\int_0^\infty d\mu \,  f_{j+1}^V(\mathbf m_j, \mu; t_1)+\\
&+&\frac{1}{2}\int_0^t dt_1\, e^{-\frac{j(j-1)}{2\, V}(t-t_1)}\sum_{\ell=1}^{j}\int_0^{m_\ell}d\mu\, f_{j+1}^V(m_1,\dots, m_\ell - \mu,\dots, m_j, \mu; t_1).\nonumber
\end{eqnarray}
that, using the operator $W_j$ defined by (\ref{S6Wj}), can be rewritten as:
\begin{eqnarray}\label{Duhamel2ante}
f_j^V(t)&=& e^{-\frac{j(j-1)}{2\, V}t}f_j^V(0)+\int_0^t dt_1\, e^{-\frac{j(j-1)}{2\, V}(t-t_1)}\ W_j\,  f_{j+1}^V(t_1) \nonumber\\
&=& \sum_{n\geq 0}^{N_0-j}\int_0^{t}dt_1\dots \int_0^{t_{n-1}}dt_n\, 
e^{-\frac{(j-1)j}{2\, V}(t-t_1)}\ \, W_{j}\, e^{-\frac{j(j+1)}{2\, V}(t_1-t_2)}\ W_{j+1}\dots \nonumber\\
&&\quad\dots e^{-\frac{(j+n-2)(j+n-1)}{2\, V}(t_{n-1}-t_n)}\ W_{j+n-1}\,  e^{-\frac{(j+n-1)(j+n)}{2\, V}t_n}f_{j+n}^V(0),
\end{eqnarray}
where we set $t_{-1}\equiv 0$, $t_0\equiv t$.\\
Property  (\ref{S6T1E2}) follows from the two following facts.  The first is that the sum:
\bean
&&\sum_{n\geq 0}^{N_0-j}\int_0^{t}dt_1\dots \int_0^{t_{n-1}}dt_n\, 
\left|\left|e^{-\frac{(j-1)j}{2\, V}(t-t_1)}\ \, W_{j}\, e^{-\frac{j(j+1)}{2\, V}(t_1-t_2)}\ W_{j+1}\dots \right.\right.\nonumber\\
&&\left.\left.\quad\dots e^{-\frac{(j+n-2)(j+n-1)}{2\, V}(t_{n-1}-t_n)}\ W_{j+n-1}\,  e^{-\frac{(j+n-1)(j+n)}{2\, V}t_n}f_{j+n}^V(0)\right|\right|_{L^1((\R^+)^j)}
\eean
is dominated by a convergent series uniformly for $t\in [0, 1/6\rho_0)$. The second is  that for every  $j\ge 1$, every $n\ge 1$:
\bear
\label{S6Convdom}
\lim_{\substack{V,\, N_0 \to +\infty\\ \frac{N_0}{V}\to \rho_0}}\left(\int d\mathbf{t}_ne^{-\frac{j(j-1)}{2V}(t-t_1)}W_{j}
\dots  e^{-\frac{(j+n-2)(j+n-1)}{2V}(t_{n-1}-t_n)}W_{j+n-1}e^{-\frac{(j+n-1)(j+n)}{2\, V}t_n}f_{j+n}^{V}(0)\right) \nonumber \\
=\rho_0^{j+n}
\frac{t^n}{n!}W_j\circ W_{j+1}\cdots \circ W_{j+n-1}[f_0^{\otimes {j+n}}]
\eear
in $L^1((\R^+)^j)$, uniformly for $t\in [0,  T]$ for any $T>0$. \\

We start proving the first. 
By (\ref{scaling2}) and (\ref{S6Wjnorma})  we obtain:
\bear
\label{S6EN1new}
&&\left\|W_{j}\circ \cdots \circ W_{j+n-1}[f_{j+n}^{V}(0)]\right\|_{L^1((\R^+)^{j})} \le \left(\frac{3}{2}\right)^n  j(j+1)
\dots (j+n-1)||f_{j+n}^V(0)||_{L^1((\R^+)^j)} \nonumber\\
&&\le\left(\frac{3}{2}\right)^n  j(j+1)\dots (j+n-1)\left(\frac{N_0}{V}\right)^{j+n}.
\eear
Since in the expression  (\ref{Duhamel2ante}) $t_n<t_{n-1}<\dots < t_1<t_0\equiv t$, it follows that $t_{k-1}-t_k>0$ for any $k=1, \dots, n$ and by (\ref{S6EN1new}) we deduce:
\begin{eqnarray}
\label{geo3}
&& 
\int d\mathbf{t}_n\,\left\|e^{-\frac{j(j-1)}{2V}(t-t_1)}W_{j}
\dots  e^{-\frac{(j+n-2)(j+n-1)}{2V}(t_{n-1}-t_n)}W_{j+n-1}e^{-\frac{(j+n-1)(j+n)}{2\, V}t_n}f_{j+n}^{V}(0)\right\|_{L^1((\R^+)^{j})}\nonumber\\
&&=\int d\mathbf{t}_n\, e^{-\frac{j(j-1)}{2V}(t-t_1)} \dots  e^{-\frac{(j+n-2)(j+n-1)}{2V}(t_{n-1}-t_n)} e^{-\frac{(j+n-1)(j+n)}{2\, V}t_n}\times \nonumber\\
&&\hskip 8cm  
\times \left\|W_{j}\circ \cdots \circ W_{j+n-1}[f_{j+n}^{V}(0)]\right\|_{L^1((\R^+)^{j})} \nonumber\\
&&\le \frac{t^n}{n!}\, \left(\frac{3}{2}\right)^n \, j(j+1)\dots (j+n-1)\ \left(\frac{N_0}{V}\right)^{j+n}.
\eear

Using again (\ref{S3E101}) and the  hypothesis (\ref{S5thermod}) we deduce, for all $j\ge 1$, $n\ge 0$ and $t>0$:
\bear
&&\int d\mathbf{t}_n\,\left\|e^{-\frac{j(j-1)}{2V}(t-t_1)}W_{j}
\dots  e^{-\frac{(j+n-2)(j+n-1)}{2V}(t_{n-1}-t_n)}W_{j+n-1}e^{-\frac{(j+n-1)(j+n)}{2\, V}t_n}f_{j+n}^{V}(0)\right\|_{L^1((\R^+)^{j})}\nonumber \\
&&\hskip 12cm \le  4^{j-1} (6\rho_0 t)^n. \label{S6EM2}
\eear
Estimate (\ref{S6EM2}) shows that  the sum giving  $f_j^V(t)$ in (\ref{Duhamel2ante})  is dominated, uniformly for  $t\in [0, \tau _1)$, $\tau _1=\frac{1}{6\rho_0}$,  by a convergent series. \\
In order to deduce (\ref{S6Convdom})  for every $j\ge 1$ and every $n\ge 1$ uniformly for $t\in [0,  \tau _1)$
we notice first of all that, as $V\to +\infty$:
\bear
\label{S6Conv}
\int d\mathbf{t}_ne^{-\frac{j(j-1)}{2V}(t-t_1)}\cdots e^{-\frac{(j+n-2)(j+n-1)}{2V}(t_{n-1}-t_n)}e^{-\frac{(j+n-1)(j+n)}{2\, V}t_n}
\longrightarrow \int d\mathbf{t}_n \equiv \frac{t^n}{n!}
\eear
uniformly for $t$ in any compact of  $[0, +\infty)$.  On the other hand by (\ref{scaling6}) and the continuity of the operators $W_k$ we have
\bear
\label{S6convW}
\lim_{\substack{N_0,\,V\to +\infty\\ \frac{N_0}{V}\to \rho_0}} \left\|W_{j}\circ \cdots \circ W_{j+n-1}\left[f_{j+n}^{V}(0)-\rho_0^{j+n} f_0^{\otimes{ j+n}}\right]\right\|_{L^1((\R^+)^{j})}=0
\eear
From  (\ref{S6Conv}) and (\ref{S6convW}),  (\ref{S6Convdom}) follows  and therefore: 
\bean
f_j^V(t)\to \sum_{n\ge 0} \frac{t^n}{n!}\rho_0^{j+n} W_{j}\circ \cdots \circ W_{j+n-1}[f_0^{\otimes{j+n}}]\equiv g_j(t)
\eean
in $\mathbf C([0, \tau _1); L^1(\R^+))\cap  \mathbf C^1((0,  \tau _1); L^1(\R^+))$. Since $f_j^V(t)\ge 0$ for all $j\in \N^*$ and $t\ge 0$ it follows that $g_j(t)\ge 0$ for $t\in [0, \tau _1)$ and all $j$.

A straightforward computation shows that  $\{g_j\}_{j\in N^*}$ satisfies system (\ref{BBGKYjS}) for $t\in[0, \tau _1)$ and this proves the Lemma  \ref{S6T1} for $T\in [0, \tau _1)$. 

In order to extend this result for $T>\tau _1$ we first notice that for all $t>0$:
\begin{eqnarray}\label{Duhamel2ante10}
f_j^V\left(t+\frac {\tau _1} {2}\right)&=& e^{-\frac{j(j-1)}{2\, V}t}f_j^V(\tau_1/2)+\int_0^t dt_1\, e^{-\frac{j(j-1)}{2\, V}(t-t_1)}\ W_j\,  f_{j+1}^V\left(t_1+\frac {\tau _1} {2}\right) \nonumber\\
&=& \sum_{n\geq 0}^{N_0-j}\int_0^{t}dt_1\dots \int_0^{t_{n-1}}dt_n\, 
e^{-\frac{(j-1)j}{2\, V}(t-t_1)}\ \, W_{j}\, e^{-\frac{j(j+1)}{2\, V}(t_1-t_2)}\ W_{j+1}\dots \nonumber\\
&&\quad\dots e^{-\frac{(j+n-2)(j+n-1)}{2\, V}(t_{n-1}-t_n)}\ W_{j+n-1}\,  e^{-\frac{(j+n-1)(j+n)}{2\, V}t_n}f_{j+n}^V(\tau_1/2),
\end{eqnarray}
In order to pass to  the limit as $V\to +\infty$ and $N_0/V\to \rho _0$ in (\ref{Duhamel2ante10}) we use the same two arguments as for  (\ref{Duhamel2ante}). More precisely, using Lemma \ref{S4T1}, we obtain:
\bear
\label{S6EN1newnew}
&&\left\|W_{j}\circ \cdots \circ W_{j+n-1}[f_{j+n}^{V}(\tau_1/2)]\right\|_{L^1((\R^+)^{j})} \le \left(\frac{3}{2}\right)^n  j(j+1)
\dots (j+n-1)||f_{j+n}^V(0)||_{L^1((\R^+)^j)} \nonumber\\
&&\le\left(\frac{3}{2}\right)^n  j(j+1)\dots (j+n-1)\left(\frac{N_0}{V}\right)^{j+n}.
\eear
and then, arguing as before:
\bear
&&\hskip -1.5cm \int d\mathbf{t}_n\,\left\|e^{-\frac{j(j-1)}{2V}(t-t_1)}W_{j}
\dots  e^{-\frac{(j+n-2)(j+n-1)}{2V}(t_{n-1}-t_n)}W_{j+n-1}e^{-\frac{(j+n-1)(j+n)}{2\, V}t_n}f_{j+n}^{V}(\tau_1/2)\right\|_{L^1((\R^+)^{j})}\nonumber \\
&&\hskip 11.5cm \le  4^{j-1} (6\rho_0 t)^n. \label{S6EM200}
\eear
On the other hand, by the continuity of the operators $W_k$ and the convergence (\ref{S6T1E2}) that we have just proved for $T\in [0, \tau _1)$ :
\bear
\label{S6convWW}
\lim_{\substack{N_0,\,V\to +\infty\\ \frac{N_0}{V}\to \rho_0}} \left\|W_{j}\circ \cdots \circ W_{j+n-1}\left[f_{j+n}^{V}(\tau_1/2)-g_j(\tau_1/2)\right]\right\|_{L^1((\R^+)^{j})}=0
\eear
From (\ref{S6EM200}) and  (\ref{S6convWW}) we deduce that, as $V\to \infty$, $N_0/V\to \rho _0$ 
\bean
&&f_j^V\left(t+\frac {\tau _1} {2}\right)\to h_j(t)\quad \hbox{in}\,\,\,\mathbf C([0, \tau _1); L^1(\R^+))\cap  \mathbf C^1((0,  \tau _1); L^1(\R^+))\\
&&h_j(t)=\sum_{n\ge 0} \frac{t^n}{n!}\rho_0^{j+n} W_{j}\circ \cdots \circ W_{j+n-1}[g_j(\tau_1/2)]\ge 0\quad \forall t\in [0, \tau _1).
\eean
We define now the function  $g_j$ for $t\in [\tau _1, 3\tau _1/2)$ as :
\bean
g_j(t)=h_j(t-\tau _1/2), \,\,\forall t\in [\tau _1, 3\tau _1/2).
\eean
Since $\{g_j(t)\}_{j\in N^*}$ satisfies system  (\ref{BBGKYjS}) for $t\in[0, \tau _1)$ it is such that:
 \bean
g\left(t+\frac {\tau _1} {2}\right)=\sum_{n\ge 0} \frac{t^n}{n!}\rho_0^{j+n} W_{j}\circ \cdots \circ W_{j+n-1}[g_j(\tau_1/2)]
\quad \forall t\in [0, \tau _1/2)
 \eean
It then follows $h_j(t)=g_j(t+\tau _1/2)$ for all $t\in [0, \tau _1/2)$.  We deduce that $g_j\in \mathbf C([0, 3\tau _1/2); L^1(\R^+))\cap  \mathbf C^1((0,  3\tau _1/2); L^1(\R^+))$ and
 \bean
&& f_j^V(t)\to g_j(t)  \quad \hbox{in}\,\,\,\mathbf C([0, 3\tau _1/2); L^1(\R^+))\cap  \mathbf C^1((0,  3\tau _1/2); L^1(\R^+))\\
&& \hbox{as}\,\,\,N_0\to \infty, \,V\to \infty,\,\,\frac {N_0} {V}\to \rho _0.
 \eean
 Since, as a plain calculation shows again, the sequence $\{h_j(t)\}_{j\in \N^*}$ satisfies system  (\ref{BBGKYjS}) for $t\in[0, \tau _1)$ and $h_j(0)=g_j(\tau _1/2)$ the sequence
$\left\{g_j\right\}_{j\in \N^*}$  satisfies system  (\ref{BBGKYjS}) for $t\in[0, 3\tau _1/2)$.

We have extended in that way the previous result on $[0, \tau _1)$ to the interval $[0, 3\tau _1/2)$. This procedure may be repeated to obtain the result in all finite interval $[0, T)$ from where Lemma \ref{S6T1} follows.
\qed 
\vskip 0.5cm 
\noindent
\textbf{Proof of Theorem \ref{MainTh}} 
From Lemma 5.3,  for any $T>0$, $f_j^V$ converges to $g_j$ in $\mathbf C([0, T); L^1(\R^+)^j)$ and $\{g_j\}_{j\in N^*}$ is a non negative solution of system  (\ref{BBGKYjS}) with initial data $\{\rho _0^j f_0^{\otimes{j}}\}_{j\in \N^*}$. By Lemma 5.2 we have $g_j=f_j^\infty$ for all $j\ge 1$ and Theorem \ref{MainTh} follows. \qed
\begin{rem} We have considered a factorized initial data for the finite system of particles (cf. (\ref{PNindat0})). We could also have considered a more general  initial data of the form:
$$
P_{N_0}(\mathbf{m}_{N_0},0)=(N_0)!\ \Psi_{N_0}(\mathbf{m}_{N_0}),\ \ \ P_N(\mathbf{m}_{N},0)=0\ \ for\ \ N\neq N_0
$$
with 
\begin{eqnarray}\label{PcorrINT}
\Psi_{N_0}(\mathbf{m}_{N_0})\geq 0,\ \ \ \ \int d\mathbf{m}_{N_0}\ \Psi_{N_0}(\mathbf{m}_{N_0})=1,
\end{eqnarray}
Then, for any $j=1,\dots,N_0$, the correlation functions at time $t=0$ would have been
\begin{eqnarray}
f_j(\mathbf{m}_{j}, 0)=\sum_{N=j}^\infty \ \frac{1}{(N-j)!}\ \int dm_{j+1}\dots \int dm_{N} P_N(\mathbf{m}_{N},0)
= \frac{(N_0)!}{(N_0-j)!}\ \ F_j^{(N_0)}(\mathbf{m}_{j}),
\end{eqnarray}
where $F_j^{(N_0)}(\mathbf{m}_{j})$ is the $j$-particle marginal distribution (in the sense of probability measures) of $P_{N_0}(\mathbf{m}_{N_0})$, namely:
$$
F_j^{(N_0)}(\mathbf{m}_{j}):=\int dm_{j+1}\dots \int dm_{N_0} \Psi_{N_0}(\mathbf{m}_{N_0})
$$
and, by definition (see (\ref{PcorrINT})), we have:
$$
\int d\mathbf{m}_{j}\ F_j^{(N_0)}(\mathbf{m}_{j})=\int d\mathbf{m}_{j}\int dm_{j+1}\dots \int dm_{N_0} \Psi_{N_0}(\mathbf{m}_{N_0})=1.
$$
Then, the rescaled correlation functions at time $t=0$ would have satisfied
\begin{eqnarray}\label{rescCORR}
\lim_{\substack{V,\, N_0 \to +\infty\\ \frac{N_0}{V}\to \rho_0}} ||f_j^V(0)||_{L^1((\R^+)^j)}=\rho _0^j
\end{eqnarray}
Now, assuming
\begin{eqnarray}\label{propOFchaos1}
\lim_{N_0\to +\infty}||F_j^{(N_0)}(0)-f_0^{\otimes j}||_{L^1((\R^+)^j)}=0
\end{eqnarray}
for some $f_0(m)$ such that
$$
f_0(m)\geq 0, \ \ \ \int dm \ f_0(m)=1,
$$
we could have proved exactly the same results that we got for the a priori factorized case.\\
Choosing an intial datum as in (\ref{propOFchaos1}) would have meant to assume propagation of chaos to hold at time $t=0$. On the other hand, what we did has been to assume hypotheses of molecular chaos to hold at time $t=0$. \\
\end{rem}
\textit{Acknowledgements.} ME is supported by Grants MTM2008-03541 and IT-305-07. FP is supported by Grant MTM 2007-62186 and  IT-305-07. 

\nocite*
\bibliographystyle{siam}
\bibliography{coagbbgky.bib}
\end{document}